 \documentclass[final,5p,times,twocolumn,numbers]{elsarticle}

\usepackage{amssymb}
\usepackage{lipsum}
\usepackage{url}
\usepackage{hyperref}
\usepackage{booktabs}
\usepackage{xcolor}
\usepackage{multirow}
\usepackage{listings}
\usepackage{xcolor}
\usepackage{subcaption}
\usepackage{acro}       
\def\BibTeX{{\rm B\kern-.05em{\sc i\kern-.025em b}\kern-.08em
    T\kern-.1667em\lower.7ex\hbox{E}\kern-.125emX}}

\definecolor{codegreen}{rgb}{0,0.6,0}
\definecolor{codegray}{rgb}{0.5,0.5,0.5}
\definecolor{codepurple}{rgb}{0.58,0,0.82}
\definecolor{backcolour}{rgb}{0.95,0.95,0.92}
\definecolor{lcol}{HTML}{AF3235}

\hypersetup{
    colorlinks=true,
    allcolors=lcol,
    linkcolor=black
}

\newcommand{\VulRepair}{\textit{VulRepair}}

\journal{Astronomy $\&$ Computing}

\newcommand{\websiteurl}{https://sites.google.com/view/llm4cve}

\lstdefinestyle{mystyle}{
    language=c,
    backgroundcolor=\color{backcolour},   
    commentstyle=\color{codegreen},
    keywordstyle=\color{magenta},
    numberstyle=\tiny\color{codegray},
    stringstyle=\color{codepurple},
    basicstyle=\ttfamily\footnotesize,
    breakatwhitespace=false,         
    breaklines=true,                 
    captionpos=b,                    
    keepspaces=true,                 
    numbers=left,                    
    numbersep=5pt,                  
    showspaces=false,                
    showstringspaces=false,
    showtabs=false,                  
    tabsize=2
}

\begin{document}

\begin{frontmatter}




\title{\textit{Found, but Not Fixed} \\ From Software Vulnerability Discovery to Fixes with Iterative LLM Pipelines}


\author[first]{Youssef Gamal}
\author[first]{Mohamad Fakih}
\author[first]{Halima Bouzidi}
\author[first]{Rahul Dharmaji}
\author[second]{Gustavo Quiros Araya}
\author[second]{Oluwatosin Ogundare} 
\author[third]{\hspace{1.5cm} Mst Ayesha Siddika}
\author[first]{Mohammad Abdullah Al Faruque}

\affiliation[first]{organization={University of California, Irvine}, department={Dept. of Electrical Engineering and Computer Science},
            city={Irvine},
            state={CA},
            country={USA}}

\affiliation[second]{organization={Siemens Technology}, 
            city={Princeton},
            state={NJ},
            country={USA}}

\affiliation[third]{organization={Jahangirnagar University}, department={Dept. of Computer Science and Engineering},
            city={Dhaka},
            country={Bangladesh}}

\begin{abstract}
Software vulnerabilities persist as a critical threat despite the deployment of sophisticated detection tools and AI assistants. While detection has advanced, the subsequent code repair phase remains a primary bottleneck; manual intervention is inherently slow, costly, and prone to introducing secondary regressions and issues. In this paper, we investigate whether an iterative, LLM-based pipeline can provide a robust solution for the automated repair of vulnerable functions in real-world code. We propose a generalized framework that leverages the reasoning capabilities of Large Language Models (LLMs) through a multi-stage feedback loop. We benchmark this approach against state-of-the-art models, including GPT-4o, GPT-5.1, and Gemini 3 flash demonstrating that such a pipeline can achieve a high quality code repair. To support the broader security community, we provide our evaluation framework, fine-tuned weights, and comprehensive experimental dataset, on our website: \url{https://sites.google.com/view/llm4cve}
\end{abstract}




\end{frontmatter}




\section{Introduction}
\label{introduction}

Software is inherently human, and as a result, it is inherently flawed. Despite decades of advancement in software engineering, human developers continue to introduce costly vulnerabilities stemming from a fundamental disconnect with cybersecurity principles~\cite{votipkadev}. These flaws serve as open invitations for adversaries seeking to exfiltrate sensitive credentials and high-value assets~\cite{changproduction}. With the frequency of exploitation on a steady trajectory upward\footnote{\href{https://www.ic3.gov/Media/PDF/AnnualReport/2023_IC3Report.pdf}{Internet Crime Complaint Center 2023 Internet Crime Report}, 2023}, the ability to bridge the gap between discovery and repair has never been more critical. The economic repercussions are staggering: billions of dollars in lost value~\cite{bugecon1, bugecon2}, millions of man-hours lost to manual resolution\footnote{\href{https://go.malwarebytes.com/rs/805-USG-300/images/Second\%20Annual\%20State\%20of\%20Ransomware\%20Report\%20-\%20Australia.pdf}{Osterman Research: Second Annual State of Ransomware Report - Australia}, 2023}, and the pervasive leakage of sensitive user data~\cite{bugimpact1, bugimpact2}. The 2021 Log4shell vulnerability ~\cite{log4shell} in the open-source Apache Log4j library remains a haunting testament to this reality, where a single exploited vulnerability in a ubiquitous library could compromise millions of systems within days, highlighting the critical importance of rapid patching.

The stakes of the \textit{program repair gap} are no longer confined to digital servers; they have become kinetic. The Internet of Things (IoT) revolution has embedded software into the physical fabric of our lives~\cite{embsysforecast}, yet these systems remain dangerously exposed~\cite{kimembsysbugs,chenembsysbugs,davidsonembsysbugs}. Exploiting these flaws can turn household devices into botnets~\cite{iotbotnet} or, more catastrophically, disrupt national power grids~\cite{iotpowergrid}. This danger is perhaps most acute in Autonomous Vehicles (AVs)~\cite{avbugs}. In safety-critical control software, a bug is not just a glitch; it is a life-threatening failure for passengers and pedestrians alike. While automated testing suites have improved our ability to verify these systems~\cite{avbugs2}, AVs remain susceptible to the same legacy software vulnerabilities that affected traditional computer systems.

The common denominator in these high-stakes systems is their profound \textit{reliance on open-source software}, predominantly written in C~\cite{opensource}. Foundational projects such as the Linux kernel\footnote{\href{https://www.kernel.org/}{Linux Kernel v6.8}, Linux Foundation, 2024}, OpenSSL\footnote{\href{https://www.openssl.org/}{OpenSSL v3.2.1}, The OpenSSL Project, 2024}, and FFmpeg\footnote{\href{https://www.ffmpeg.org/}{FFmpeg v6.1.1}, FFmpeg Team, 2023} are the bedrock of modern infrastructure and, consequently, the primary targets for global adversaries. This has resulted in a disproportionate accumulation of Common Vulnerabilities and Exposures (CVEs) within these codebases~\cite{manycve}. While the CVE system provides a vital centralized repository for documenting vulnerabilities, it exposes a systemic weakness: CVE entry is a diagnosis, not a fix. Knowing a vulnerability exists within software code does not facilitate its repair.

\begin{figure}[!t]
    \centering
    \includegraphics[width=.8\linewidth]{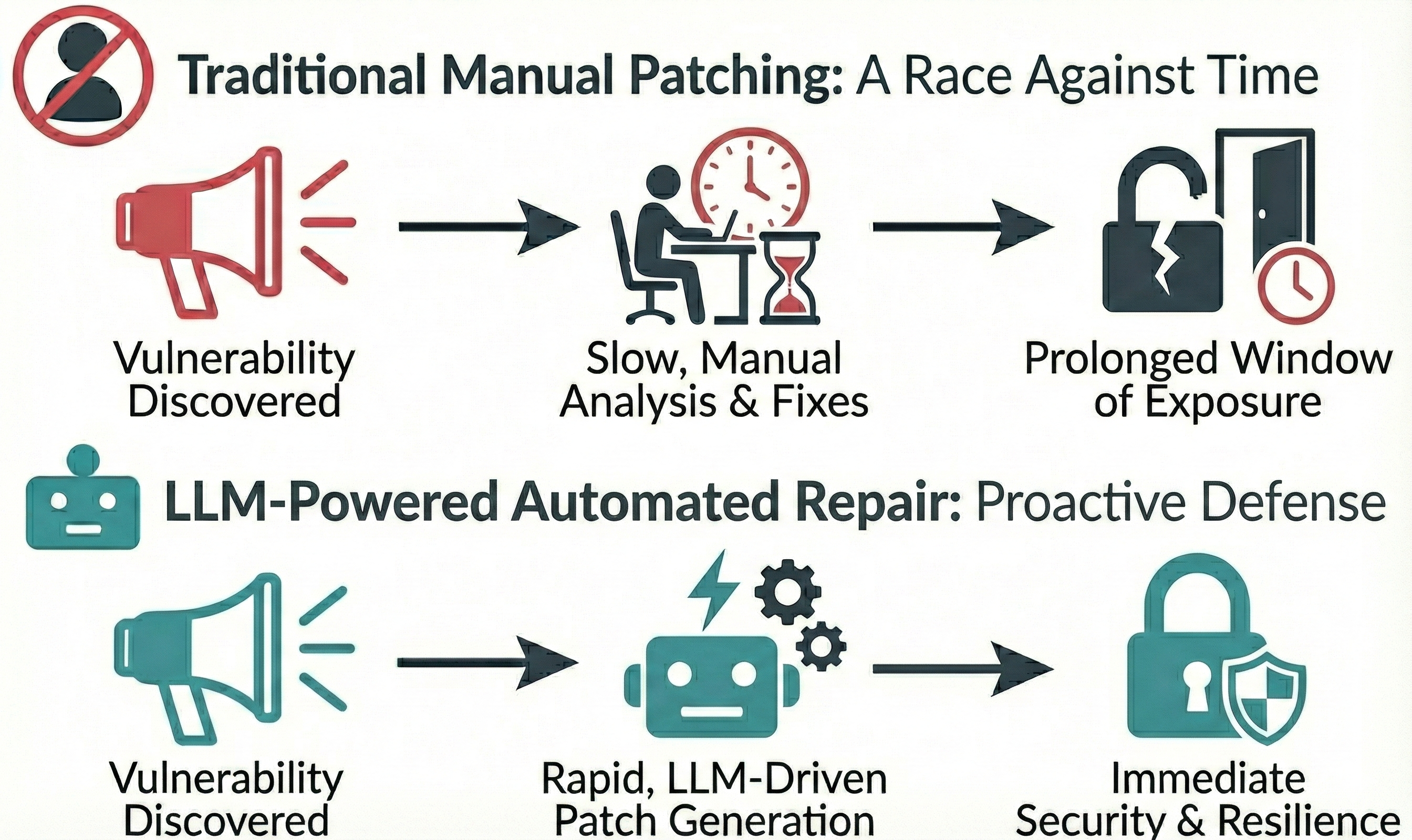}
    \caption{Bridging the gap: How an iterative repair pipeline prevents exploitation by shortening the time-to-repair for software code.}
    \label{fig:timeline}
\vspace{-0.2cm}
\end{figure}

Existing Automated Program Repair (APR) methods strive to patch software with minimal human intervention~\cite{pinvulnrev, zhangvulnrev, furepair, adamvulnrev, fuvision, harergan}. However, these systems often hit a \textit{complexity ceiling}. Traditional automated repairs frequently generate invalid code~\cite{avrproblem1} or introduce secondary security regressions~\cite{avrproblem2}. Most critically, these techniques typically require a subject-matter expert to guide the implementation~\cite{avrproblem3}. This introduces a critical dependency: when vulnerabilities are found in unmaintained legacy code or when expert intervention is unavailable, the time-to-fix increases dramatically, creating an extended window of exposure for attackers.

The emergence of highly capable Large Language Models (LLMs) represents a paradigm shift in addressing this vulnerability repair bottleneck. While early LLM implementations were prone to generating flawed or uncompilable code~\cite{liucorrectness}, state-of-the-art architectures like GPT-5.1\footnote{\href{https://openai.com/index/gpt-5-1/}{GPT-5}, OpenAI, 2025}, Gemini 3 flash\footnote{\href{https://deepmind.google/models/gemini/flash/}{Gemini 3 Flash}, Google DeepMind, 2025}, and Llama 3\footnote{\href{https://llama.meta.com/llama3/}{Llama 3}, Meta Research, 2024} have demonstrated sophisticated reasoning and code synthesis capabilities. The rise of specialized models such as Code Llama~\cite{codellama}, alongside Parameter-Efficient Fine-Tuning (PEFT)~\cite{peft} and Low-Rank Adaptation (LoRA)~\cite{lora}, has made it possible to tailor these models for the specific constraints of security patching~\cite{liu2022peft, xiongfinetune}. Furthermore, innovations in Mixture-of-Experts (MoE)~\cite{mistralmoe} and the evolution of Prompt Engineering~\cite{prompteng1, prompteng2, prompteng3} as well as sophisticated pipelines~\cite{LLM4PLC,loop1, loop2} provide a robust toolkit for refining the accuracy and relevance of AI-generated repairs.

In this work, we move beyond one-shot code generation to propose an iterative, LLM-driven repair paradigm, as illustrated in Figure~\ref{fig:timeline}. Rather than treating repair as a single predictive task, our approach treats it as a continuous refinement loop. We introduce LLM4CVE, a concrete implementation of this iterative pipeline that integrates LLMs with real-world CVE data to synthesize repair patches with minimal human oversight. Given a vulnerable code snippet, the pipeline iteratively generates, evaluates, and refines candidate fixes until a viable solution is reached. By incorporating LoRA fine-tuning and optimized prompt engineering, our framework automates the vulnerability repair process for complex, real-world CVEs. Our primary contributions include:

\begin{itemize}
    \item A novel automated pipeline for fixing security vulnerabilities in real-world software programs that requires minimal intervention from a skilled domain expert.
    \item An optimized automated and iterative process that leverages LLMs reasoning capabilities to refine its output, thereby improving upon conventional tools.
    \item A detailed evaluation of LLM4CVE for repairing various classes of CVEs and CWEs using LLMs with a diverse range of capabilities---spanning frontier proprietary systems to state-of-the-art open-source architectures---including Llama 3, GPT-3.5-Turbo\footnote{\href{https://developers.openai.com/api/docs/models/gpt-3.5-turbo}{GPT-3.5 Turbo}, OpenAI, 2026}, GPT-4o\footnote{\href{https://openai.com/index/hello-gpt-4o/}{GPT-4o}, OpenAI, 2024}, GPT-5.1, and Gemini 3 flash. This cross-section of models serves to showcase the performance and versatility of the LLM4CVE framework across different training philosophies and parameter scales.
\end{itemize}

\section{Background}
\label{sec:bg}

\subsection{The Taxonomy of Failure: CVEs \& CWEs}
\label{sec:bgcve}

The modern security landscape relies heavily on centralized repositories to document and track software flaws. Entities such as the National Vulnerability Database (NVD)\footnote{\href{https://nvd.nist.gov/}{NVD}, National Vulnerability Database, 2024} and MITRE’s Common Vulnerabilities and Exposures (CVE)\footnote{\href{https://cve.mitre.org/}{CVE}, MITRE Corporation, 2026} serve as the industry standard for reporting discovered bugs. When a vulnerability is identified in widely-deployed software, it is assigned a unique CVE identifier, used for sharing technical descriptions and mitigation strategies~\cite{changproduction}. 

While CVEs target specific, instance-based software flaws, the Common Weakness Enumeration (CWE) provides a formal language for describing underlying vulnerability types (e.g., Buffer Overflows, SQL Injection)~\cite{householdercwe}. However, a critical limitation exists: these databases are designed for identification, not rectification. They provide a comprehensive diagnosis of a system's failure but offer no automated mechanism for the repair. Bridging this gap requires transitioning from manual documentation to an automated repair paradigm that can interpret these complex reports and synthesize functional code.

\subsection{The Lifecycle of Vulnerability Repair}
The ultimate objective of any repair technique is to remediate software deficiencies, faults, errors, or failures~\cite{bugconcepts}, before they can be exploited. This process, often referred to as APR, typically follows a three-phase lifecycle: (1) \textbf{Detection}, (2) \textbf{Patching}, and (3) \textbf{Verification}~\cite{pinvulnrev}.

\begin{enumerate}
    \item \textbf{Detection:} Identifying the precise location of a flaw using static analysis~\cite{zhengstaticdetect}, machine learning classifiers~\cite{limldetect, chakramldetect}, or dynamic fuzzing~\cite{sotirovdetect}.
    \item \textbf{Patching:} Generating a candidate fix. Traditional APR relies on heuristic-based or symbolic execution methods~\cite{monvulnrev}, which often struggle with the semantic complexity of real-world code. Deep learning approaches~\cite{zhangvulnrev} have begun to address this by learning from massive datasets of human-authored commits.
    \item \textbf{Verification:} Ensuring the generated patch is both functional and secure. This is typically achieved through regression test suites~\cite{xiallmrepair3, avrproblem1}, formal verification, or expert manual review~\cite{gaorepair}.
\end{enumerate}

\begin{figure}[ht!]
    \centering
    \includegraphics[width=.75\linewidth]{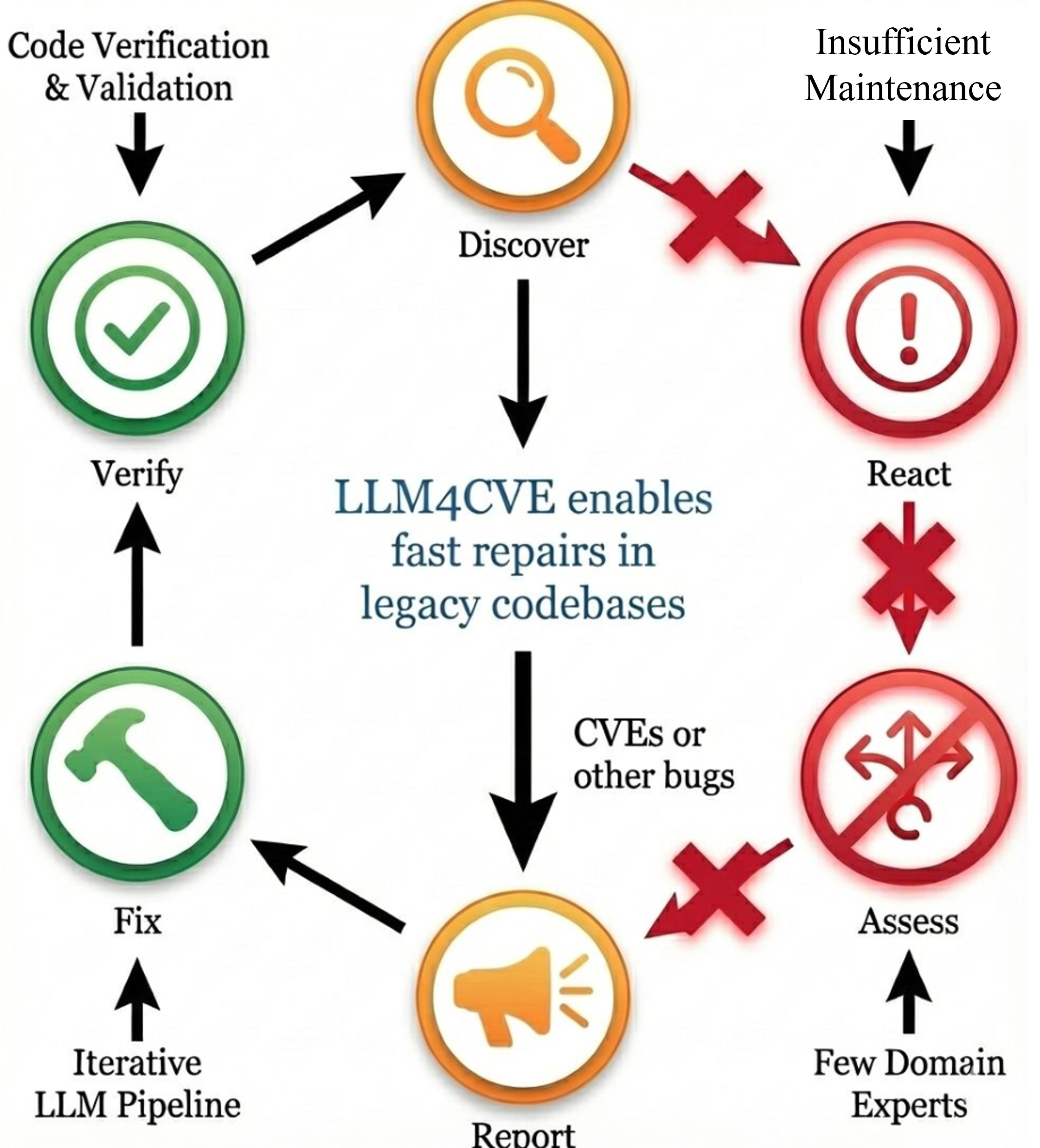}
    \caption{The software vulnerability detection-and-repair lifecycle. Our iterative approach augments the traditional cycle by shortening the time-to-repair through autonomous feedback loops.}
    \label{fig:vcycle}
\vspace{-0.5cm}
\end{figure}

Despite these established phases, traditional APR remains brittle, often producing patches that pass tests but fail in production or introduce new vulnerabilities. Our approach seeks to augment the first two stages of this lifecycle, as shown in Figure~\ref{fig:vcycle}, by introducing a feedback-driven, iterative loop that mimics the way a human developer refines code.

\subsection{Large Language Models in Software Security}

The rise of LLMs has fundamentally altered the APR landscape, offering a potential solution to the \textit{repair gap} by transforming diagnostic data into functional code. Based on the Transformer architecture and multi-head attention mechanisms~\cite{attention}, LLMs excel at capturing long-range dependencies and the complex syntax of programming languages~\cite{shillmcontext, zaheercontext}. Unlike traditional rule-based APR tools, the attention mechanism allows these models to maintain context across large source files, a critical requirement for fixing vulnerabilities where the cause (e.g., an unvalidated input) and the effect (e.g., a buffer overflow) may be spatially distant within the codebase.
While early models demonstrated surprising emergent coding abilities~\cite{llmemerge}, the field has shifted toward highly specialized architectures designed for deep reasoning. Proprietary models like OpenAI's GPT-4o offer immense reasoning power but remain \textit{black boxes} with restrictive APIs\footnote{\href{https://openai.com/api/pricing}{OpenAI API Pricing}, OpenAI, 2026}, which can limit the depth of security-specific analysis and iterative refinement. Consequently, the research community has pivoted toward high-performance, open-source alternatives like \textit{Llama 3} and \textit{Mixtral}~\cite{mistralmoe}. 
These open-source models provide the deep transparency required for the domain-specific fine-tuning (e.g., on CVE/CWE datasets) central to our study. This shift directly addresses our central research question: \textit{Can an iterative pipeline finally bridge the vulnerability repair gap?} By leveraging the inherent reasoning capabilities of these models and adapting them to the specific logic of vulnerability repair, we can move beyond merely '\textit{finding}' bugs to autonomously '\textit{fixing}' them through a feedback-driven, iterative process.

\begin{figure}[ht!]
    \centering
    \includegraphics[width=0.98\linewidth]{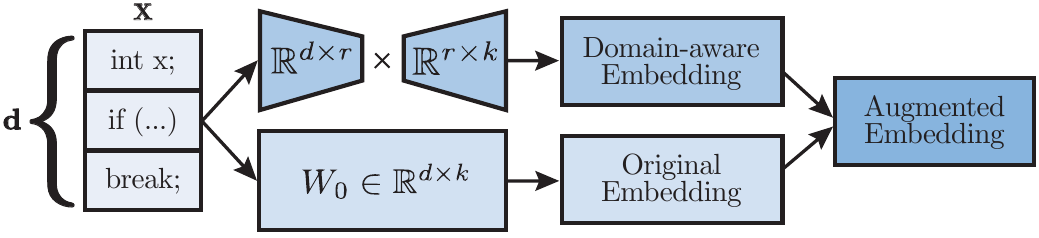}
    \caption{An Illustration of the Low-Rank Adaptation (LoRA) Process.}
    \label{fig:lora}
\vspace{-0.5cm}
\end{figure}

\subsubsection{Parameter-Efficient Augmentation}
To bridge the gap between general-purpose reasoning and specialized security repair, models must be adapted to the nuances of specific CWEs. PEFT, specifically LoRA~\cite{lora} (See Fig. \ref{fig:lora}), has emerged as the gold standard for this task. The primary challenge in adapting LLMs for security is their immense parameter count, which makes full-parameter fine-tuning computationally prohibitive and prone to catastrophic forgetting of general reasoning capabilities.

LoRA circumvents this by freezing the pre-trained model weights $W_0 \in \mathbb{R}^{d \times k}$ and injecting trainable rank decomposition matrices. For a given weight update, the new weight $W$ is represented as:
\begin{equation}
    W = W_0 + \Delta W = W_0 + BA
\label{eq:fine_tune}
\end{equation}
where $B \in \mathbb{R}^{d \times r}$, $A \in \mathbb{R}^{r \times k}$, and the rank $r \ll \min(d, k)$. This approach drastically reduces the number of trainable parameters, often by a factor of 10,000, while maintaining the model's ability to synthesize specialized security patches~\cite{zakenpeft, liupeft}. When combined with 4-bit quantization through \textit{QLoRA}~\cite{qlora}, these models can be fine-tuned on consumer-grade hardware to remediate complex legacy codebases that were previously inaccessible to automated tools.

\begin{figure*}[t]
    \centering
    \includegraphics[width=.9\textwidth]{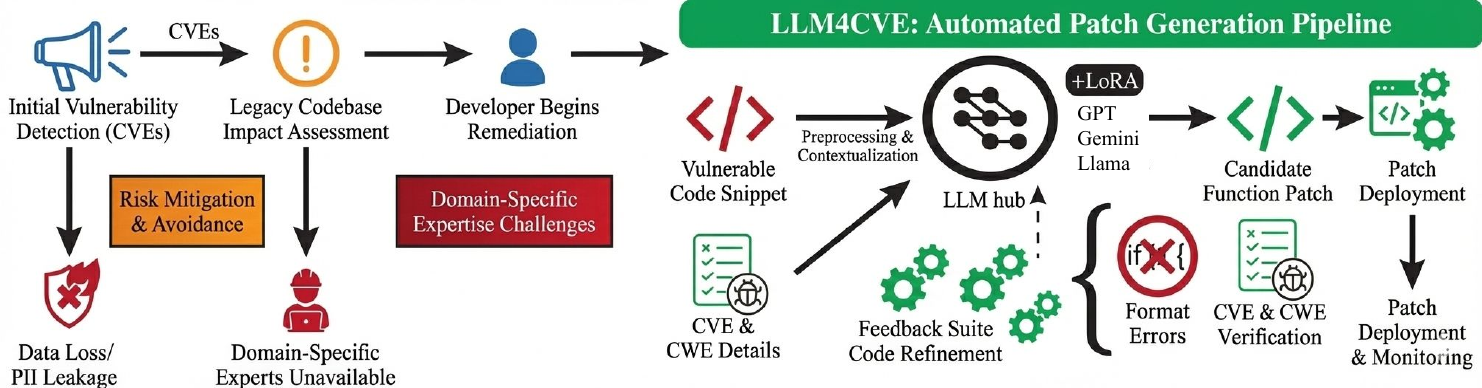}
    \caption{A visualization of how the LLM4CVE pipeline can automatically fix common software vulnerabilities}
    \label{fig:pipeline_overview}
    \vspace{-0.5cm}
\end{figure*}

\subsubsection{Iterative Prompt Engineering \& Reasoning}
The quality of an automated repair is largely a function of the context provided to the model. \textit{Prompt Engineering} techniques, such as Chain-of-Thought (CoT)~\cite{chainofthought}, enable models to decompose a complex vulnerability, like a multi-stage memory leak or a race condition, into manageable logical steps. Rather than a black-box prediction, CoT encourages the model to generate an internal rationale before synthesizing the final patch, which significantly reduces the likelihood of hallucinating insecure code.
However, the \textit{Found, but Not Fixed} gap often exists because security repairs are rarely correct on the first attempt. Our pipeline moves beyond static, zero-shot prompting by implementing an iterative feedback loop. In this paradigm, the model is treated as an agent within an environment where it receives external scores from code analysis tools. If the score of a generated patch is below a certain threshold, the error logs are fed back into the model. This mimics the \textit{Think Step-by-Step}~\cite{kojimazeroshot} methodology but extends it across multiple generations, allowing the LLM to self-reflect and improve its output. This iterative reasoning allows our approach to handle real-world CVEs that possess semantic complexities far beyond the reach of standard one-shot generators.

\section{Methodology}
\label{sec:methods}

The core contribution of our work is a generalized, closed-loop framework for transforming vulnerability diagnostics into functional code. Our approach, realized through the LLM4CVE pipeline, treats APR as an iterative refinement problem rather than a one-shot predictive task. The architectural design of the pipeline, illustrated in Figure~\ref{fig:pipeline_overview}, centers on an autonomous feedback loop between semantic code analysis and LLM-driven synthesis.

\subsection{Vulnerability Selection \& Dataset Composition}
To evaluate the pipeline's robustness against diverse failure modes, we select eight of the most critical Common Weakness Enumerations (CWEs) identified by MITRE. These classes represent the primary drivers of modern software exploitation, ranging from memory corruption to improper input handling. Table~\ref{tab:cwetable} provides an overview of the selected CWEs and their prevalence within our dataset, which is derived from real-world CVE reports. Section~\ref{sec:dataset} further details the automated filtration and pre-processing steps required to ingest these vulnerabilities into the repair pipeline.

\vspace{-0.3cm}
\begin{table}[!ht]
\centering
\caption{An overview of selected CWEs for the LLM4CVE pipeline}
\scalebox{.75}{
\begin{tabular}{c|llccl} 
\cline{1-5}
\multirow[c]{2}{*}{Language} &\multirow[c]{2}{*}{CWE}     & \multirow[c]{2}{*}{Description}                    & MITRE & \multirow[c]{2}{*}{Count}  \\
& & & Link &\\
\cline{1-5}
\multirow{8}{*}{C}& CWE-125 & Out-of-bounds Read             & \href{https://cwe.mitre.org/data/definitions/125.html}{Link}       & 452              &   \\
& CWE-119 & Improper Buffer Access         & \href{https://cwe.mitre.org/data/definitions/119.html}{Link}       & 363              &   \\
& CWE-20  & Improper Input Validation      & \href{https://cwe.mitre.org/data/definitions/20.html}{Link}       & 289              &   \\
& CWE-787 & Out-of-bounds Write            & \href{https://cwe.mitre.org/data/definitions/787.html}{Link}       & 179              &   \\
& CWE-476 & NULL Pointer Dereference       & \href{https://cwe.mitre.org/data/definitions/476.html}{Link}      & 176              &   \\
& CWE-190 & Integer Overflow or Wraparound & \href{https://cwe.mitre.org/data/definitions/190.html}{Link}       & 156              &   \\
& CWE-120 & Classic Buffer Overflow        & \href{https://cwe.mitre.org/data/definitions/120.html}{Link}       & 121              &   \\
& CWE-416 & Use After Free                 & \href{https://cwe.mitre.org/data/definitions/416.html}{Link}      & 120              &   \\
\cline{1-5}
\multirow{10}{*}{Python}&CWE-79 & Cross-site Scripting             & \href{https://cwe.mitre.org/data/definitions/79.html}{Link}       & 548              &   \\
&CWE-20 & Improper Input Validation             & \href{https://cwe.mitre.org/data/definitions/20.html}{Link}       & 499               &   \\
&CWE-918 & Server-Side Request Forgery             & \href{https://cwe.mitre.org/data/definitions/918.html}{Link}       & 308               &   \\
&CWE-89 & SQL Injection             & \href{https://cwe.mitre.org/data/definitions/89.html}{Link}       & 294               &   \\
&CWE-193 &Off-by-one Error              & \href{https://cwe.mitre.org/data/definitions/193.html}{Link}       & 228               &   \\
&CWE-22 & Path Traversal            & \href{https://cwe.mitre.org/data/definitions/22.html}{Link}       & 199               &   \\
&CWE-352 & Cross-Site Request Forgery             & \href{https://cwe.mitre.org/data/definitions/352.html}{Link}       & 191               &   \\
&CWE-200 & Exposure of Sensitive Information             & \href{https://cwe.mitre.org/data/definitions/200.html}{Link}       & 175               &   \\
&CWE-601 & Open Redirect             & \href{https://cwe.mitre.org/data/definitions/601.html}{Link}       & 147               &   \\
&CWE-94 & Code injection             & \href{https://cwe.mitre.org/data/definitions/94.html}{Link}       & 141               &   \\
\cline{1-5}
\end{tabular}
}
\label{tab:cwetable}
\vspace{-0.3cm}
\end{table}

\subsection{Guided vs. Unguided Prompt Engineering}
\label{sec:prompteng}

A critical question in LLM-based repair is how much context is required to move from a diagnosis to a fix. We employ two distinct prompting strategies: \textit{guided} one-shot and \textit{unguided} zero-shot prompts. 

\begin{itemize}
    \item \textbf{Guided Prompting:} Provides the model with the specific CVE/CWE identifiers, technical descriptions of the flaw, and prescriptive instructions for repair. This mimics a scenario where a developer uses a documented vulnerability report to guide their fix.
    \item \textbf{Unguided Prompting:} Instructs the model to repair the code without providing external vulnerability details. This tests the model's inherent ability to zero-in on flaws through code reasoning alone.
\end{itemize}

To ensure the output is machine-consumable, we mandate the use of code delimiters and strictly instruct the model to produce compilable code with minimal semantic drift. This minimal change constraint is vital for maintaining the original functionality of the software.

\subsection{Semantic Code Analysis via CodeBLEU}
\label{sec:codeanalysis}

Patching the vulnerable code is mostly localized within a few lines making the fixed code semantically similar to the vulnerable one. We use CodeBLEU~\cite{codebleu} as a primary metric to measure such similarity. Unlike standard BLEU scores used in natural language, CodeBLEU incorporates Abstract Syntax Trees (ASTs) and Dataflow Graphs (DFGs) to assess whether the logical structure of the repaired code aligns with ground-truth patches. This encourages the functionality of the LLM's output code to align with that of the vulnerable one.

\subsection{The Iterative Repair Paradigm}
\label{sec:iterate}

The cornerstone of our methodology is the iterative feedback loop, which allows the pipeline to self-correct and refine its patches. Single one-shot generation prompts an LLM to fix the vulnerable code $V$ and uses the output as the fixed code patch $\hat{F}$. Rather than accepting the LLM output directly, our system runs multiple iterative loops. In loop $i$, it evaluates the similarity score $CodeBLEU(\hat{F}_{i-1}, V)$ between the input code and the previous output . If the score is below 0.75, an error message is appended to the feedback prompt instructing the LLM to regenerate a fixed code patch $\hat{F}_i$ and make it more aligned with the original vulnerable code $V$ functionality.

In this stage, the LLM is provided with the previous attempt ${\hat{F}_{i-1}}$ and the associated error feedback within the same context window. This agentic behavior allows the model to analyze why its previous patch failed and adjust its strategy, bridging the gap through active trial-and-error. The output ${\hat{F}_{i}}$ that achieves the highest similarity score $CodeBLEU({\hat{F}_{i}},V)$ is selected as the final candidate for human-verified evaluation. We cap this process at one extra iteration, thus giving the LLM a total of two iterations to generate the patched code.  While having more iterations is possible, we limit it to two to ensure that the whole prompt with the feedback fit in the context window of all of the LLMs tested.

\subsection{Real-World Evaluation Challenges}

Unlike APR research conducted on synthetic datasets, our evaluation targets real-world codebases where functions are often deeply nested and rely on external dependencies. This makes traditional static analysis and unit testing infeasible for fragmented snippets. Consequently, we focus on:
\begin{enumerate}
    \item \textbf{Semantic Fidelity:} Comparing candidate patches to historical ground-truth fixes using CodeBLEU.
    \item \textbf{Syntactic Viability:} Performing end-to-end compilation checks on synthesized functions.
    \item \textbf{Code Repair Efficacy:} Evaluating whether the patch effectively closes the specific security window, even if the implementation differs slightly from the developer's original commit.
\end{enumerate}

\section{Experimental Setup}
\label{sec:setup}

In this section, we discuss the design of our experimental apparatus and provide details on the pre-processing, testing, and evaluation schemes of our work.

The experiments are conducted using multiple compute nodes equipped with one Nvidia A100 (80GB), 48 CPU cores and 256GB of system memory.

\subsection{Datasets Preparation}
\label{sec:dataset}

The foundation of our evaluation is the C code examples in \textit{CVEFixes} ~\cite{cvefixes} dataset, and the Python code ones in \textit{MoreFixes} ~\cite{morefixes}. These comprehensive datasets contain metadata, commit histories, and CWE classifications. While these repositories provide a vast repository of security flaws, they are not inherently structured as a set of labeled, function-level repair pairs. To address our research goal of autonomous repair, we developed a specialized preprocessing pipeline to extract high-fidelity before-and-after code snippets. Our curation process follows a multi-stage filtration strategy to ensure the quality and viability of the candidate patches: 

\begin{enumerate} 

\item \textbf{CWE Filtering:} We first isolate all function-level changes organized by their unique CVE identifiers. To ensure our evaluation focuses on well-defined vulnerability classes, we exclude entries with ambiguous or missing metadata, such as \textit{NVD-CWE-noinfo} and \textit{NVD-CWE-other}. 

\item \textbf{Structural Pair Matching:} We specifically target CVEs that exhibit a one-to-one repair pattern. This involves identifying unique pairs where a single vulnerable \textit{before} snippet is directly corrected by a single non-vulnerable \textit{after} snippet within the same function scope. This structure provides the unambiguous ground truth necessary for system evaluation.

\item \textbf{Context Constraint Management:} To accommodate the context window limitations of modern LLMs, we prune the dataset to remove any code pairs where either snippet exceeds a 500-token threshold. This ensures that the model can ingest the entire vulnerable function and its associated prompt without performance degradation. 

\end{enumerate} 

Following this rigorous filtration, our final curated dataset comprises eight distinct CWE classes, as detailed in Table~\ref{tab:cwetable}. Each class contains at least 100 candidate pairs. This corpus allows us to move beyond easy examples and test the iterative pipeline against a broad spectrum of real-world logic flaws and memory corruption vulnerabilities.

\subsection{Large Language Model Selection} 
\label{sec:llms} 

To rigorously evaluate the efficacy of our iterative repair paradigm, we select a cross-section of frontier models representing both closed-source proprietary systems and state-of-the-art open-source architectures. Specifically, we employ \textbf{GPT-3.5-Turbo}, \textbf{GPT-4o}, \textbf{GPT-5.1}, \textbf{Gemini 3 flash}, \textbf{Llama 3 8B}, and \textbf{Llama 3 70B}. This selected set of LLMs allows us to investigate the performance gradient across different parameter scales and training philosophies. The inclusion of fine-tuned \textit{Llama 3} family is particularly critical for our research goal of bridging the code repair gap through specialization. Their open-weights nature permits the application of \textit{LoRA} fine-tuning (as detailed in Section~\ref{sec:augment}), enabling us to adapt the models to the specific semantic nuances of C-language security patches. In contrast, the \textit{GPT} and \textit{Gemini} models provide a high-reasoning baseline for zero-shot and one-shot capabilities within the iterative loop making them easily applicable to fixing vulnerabilities in both C and Python programming languages.

A primary consideration in our LLM selection process was the maximum context window of each model. Our iterative pipeline relies on the model’s ability to retain the original vulnerable code, the previous failed attempt, and the associated feedback tokens within a single context session. As shown in Table~\ref{tab:modelinfo}, all selected models provide a sufficiently large window to accommodate our 500-token function limit alongside iterative instructions. While GPT-4o, GPT-5.1 and Gemini 3 flash are inherently multimodal, our framework exclusively leverages their text-based reasoning to maintain parity with the other code-centric models. 

\begin{table}[ht!]
\centering
\caption{Context lengths and parameter counts of selected LLMs. 
}
\scalebox{1.}{
\begin{tabular}{ccc} 
\hline
Model       & Context Length (Tokens) \\ 
\hline
GPT-3.5-Turbo     & 16,385                              \\
GPT-4o      & 128,000                           \\
GPT-5.1 & 400,000                                \\
Gemini 3 flash & 1,000,000                                \\
Llama 3 8B  & 8,192                                  \\
Llama 3 70B & 8,192                                 \\
\hline
\end{tabular}
}
\label{tab:modelinfo}
\vspace{-0.3cm}
\end{table}

\subsection{LLM Augmentation via Low-Rank Adaptation}
\label{sec:augment}

To bridge the gap between general-purpose code reasoning and the specialized logic required for security remediation, we perform supervised fine-tuning on the \textit{Llama 3 70B} model using LoRA. LoRA operates on the principle that weight updates during domain adaptation possess a \textit{low intrinsic rank} allowing us to freeze the pre-trained weights and inject trainable rank decomposition matrices into the model layers as mathematically expressed in Eq. \ref{eq:fine_tune}. This approach allows for high-fidelity adaptation to the semantic patterns of C-language security patches while reducing the trainable parameter count by a factor of up to 10,000, thereby preventing \textit{catastrophic forgetting} of the model's reasoning abilities.

We fine-tune Llama 3 70B using LoRA implemented via the Unsloth\footnote{\href{https://unsloth.ai/}{Easily run \& train models locally}, Unsloth, 2026} library. To reduce GPU memory requirements, the model is loaded in 4-bit precision (QLoRA). LoRA adapters are applied to all attention and feed-forward projection layers with a rank of $r = 16$ and a scaling coefficient of $\alpha = 16$. The model is trained for 60 steps using the AdamW optimizer with a learning rate of $2 \times 10^{-4}$ and a batch size of 8. We utilize a 90/10 train/test split from our curated dataset of C-code (See Sec. \ref{sec:dataset}). This ratio is chosen to maximize the training corpus for the complex, low-frequency patterns found in real-world \textit{broken/fixed} pairs while reserving a statistically significant subset for unbiased evaluation. During training, the model learns to map specific vulnerability signatures, such as unchecked buffer indices, to their respective human-verified corrections. The final model is then integrated into the iterative pipeline to rectify unseen vulnerable code samples under varying levels of guidance.

\subsection{Iterative Pipeline Configurations}
\label{sec:configs}

To isolate the factors that contribute to successful code and concretely address the repair gap, we evaluate our models across three distinct pipeline configurations:

\begin{enumerate}
    \item \textbf{Unguided (Zero-Shot):} The LLM model is prompted to repair code using only its pre-trained knowledge (no fine-tuning), without external CVE/CWE context.
    \item \textbf{Guided (One-Shot):} The prompt is augmented with metadata, including CVE identifiers and CWE descriptions, to simulate a developer following a diagnostic report.
    \item \textbf{Guided + Feedback (Iterative):} This represents our full agentic workflow. The model receives iterative feedback from compiler logs and CodeBLEU metric analysis, allowing it to refine its patch over multiple generations in a guided fashion. For the \textit{Llama 3 70B} model, this configuration also incorporates the specialized LoRA adapter.
\end{enumerate}

Our evaluation suite includes \textbf{six} tested models, resulting in \textbf{18} unique model/configuration combinations. A comprehensive taxonomy of these configurations is provided in Table~\ref{tab:configtable}.

\begin{table}[ht!]
\centering
\caption{Models \& pipeline configurations used in our study. 
}
\scalebox{.85}{
\begin{tabular}{c|ccc} 
\hline
\multirow{2}{*}{Model} & \multicolumn{3}{c}{Iterative Pipeline Configuration}  \\ 
\cline{2-4}
                       & Unguided    & Guided       & Guided + Feedback         \\
\hline
GPT-3.5-Turbo                & $\checkmark$ & $\checkmark$ & $\checkmark$            \\ 
\hline
GPT-4o                 & $\checkmark$ & $\checkmark$ & $\checkmark$            \\ 
\hline
GPT-5.1                 & $\checkmark$ & $\checkmark$ & $\checkmark$            \\ 
\hline
Gemini 3 flash                 & $\checkmark$ & $\checkmark$ & $\checkmark$            \\ 
\hline
Llama 3 8B             & $\checkmark$ & $\checkmark$ & $\checkmark$            \\ 
\hline
Llama 3 70B            & $\checkmark$ & $\checkmark$ & $\checkmark$            \\
\hline
\end{tabular}
}
\label{tab:configtable}
\end{table}

\subsection{Automated Feedback Pipeline}
\label{sec:iterateimpl}

The iterative mechanism is the primary engine for bridging the repair gap by transforming a static \textit{diagnosis} into a verified \textit{fix}. The pipeline employs a \textit{guided with feedback} architecture, which integrates sophisticated prompt engineering with a recursive generation loop to synthesize high-fidelity security patches. To maintain computational efficiency and prevent the degradation of reasoning capabilities through \textit{catastrophic forgetting}, the iterative process is constrained to two cycles.

For each iteration, the vulnerable code snippet and associated prompts are limited to approximately 500 tokens to ensure the entire context, including previous candidate patches and feedback tokens, remains within the model's operational window. Upon generating an output, the pipeline extracts the candidate code and assesses its semantic integrity using the change in CodeBLEU scores. A significant divergence in this score serves as a heuristic for potential errors, which is then used to inform the model's subsequent prompt. Crucially, to maintain the realism of the evaluation, the pipeline never utilizes the ground-truth fix for feedback, as such data would be unavailable in a real-world deployment scenario. The overall flow of this iterative step is illustrated in Figure~\ref{fig:pipeline_overview}.

\subsection{Candidate Patch Extraction} 
\label{sec:patchextract}

To automate the transition from model output to deployable code, we implement a robust extraction layer. While LLMs are instructed to use specific delimiters, their raw output often includes natural language commentary or unformatted text. Our results show that approximately 95\% of responses are well-formed and easily parsed. For the remaining 5\% of edge cases, we employ specialized regex-based logic to isolate the candidate patch from surrounding noise. In rare instances where no valid code is produced (less than 1\% of samples), the pipeline records a failure to generate a candidate, ensuring the integrity of the downstream evaluation. 

\subsection{Evaluation Metrics} 
\label{sec:metrics} 

To validate whether our generalized pipeline successfully bridges the code repair gap, we evaluate our approach using a \textit{pass@k} scheme with $k=1$, employing four primary metrics: 

\begin{enumerate} 
\item \textbf{CodeBLEU Semantic Similarity:} This metric quantifies the alignment between LLM-generated fixes and ground-truth code. By analyzing Data-Flow Graphs and Abstract Syntax Trees, CodeBLEU ensures the patch is logically sound rather than just lexically similar. 

\item \textbf{End-to-End Compilation:} We rigorously test the patches by applying them to real-world codebases and executing full project builds. This validates the syntactic correctness and '\textit{compilability}' of the iterative output. 

\item \textbf{Engineering Effort Analysis:} Finally, we evaluate the reduction in manual labor by comparing setup times and required domain expertise against traditional manual code repair workflows. 

\end{enumerate}

\subsection{Comparison with SOTA pipelines}
\label{sec:CompareSOTAsetup}
To further assess the performance of our iterative pipeline, we also compare it to \VulRepair~\cite{furepair} that fine-tunes the sequence-to-sequence model CodeT5~\cite{codet5} to patch code vulnerabilities in the C-language. The \textit{VulRepair} pipeline pre-processing strips away all code styling indentations and adds special markers to highlight the start and end of vulnerable parts in the input code. Then, the model predicts the fixed version of each vulnerable part with special delimiter markers between them. To compare \textit{VulRepair} with our pipeline, we develop a script to reconstruct the complete fixed code and remove any leftover markers. In this experiment, the maximum allowed context length is limited to only 450 instead of 500. This is done to leave extra room for any extra markers added by the \textit{VulRepair} pre-processing without surpassing its context window.

\begin{figure}[ht!]
    \centering
    \includegraphics[width=0.8\linewidth]{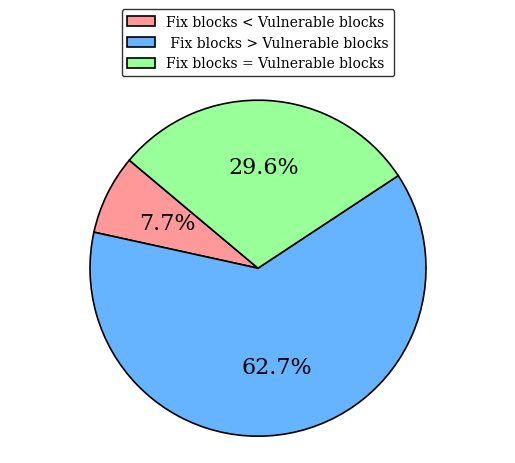}
    \caption{VulRepair vulnerable and fixed code block count mismatch ratio.}
    \label{fig:VulRepair_Errors}
\vspace{-0.3cm}
\end{figure}

However, the number of modified snippets produced by \textit{VulRepair} usually did not match the number of vulnerable snippets in the original vulnerable code as shown in Figure \ref{fig:VulRepair_Errors}. All cases where the number of modified snippets are less than that of the vulnerable snippets are reported as failures. Conversely, if the opposite is the case, the extra modified snippets are ignored.

For a fair comparison, our pipeline is also provided the marked version of the code. Additionally, clang-format \footnote{\href{https://clang.llvm.org/docs/ClangFormat.html}{Clang 23.0.0 Documentation}, Clang, 2026} is used to normalize the styling of both the vulnerable and fixed codes of both pipelines to rule-out code styling differences as a factor affecting the semantic similarity scores.

\section{Results}
\label{sec:results}

\subsection{Semantic Fidelity via CodeBLEU Scores}
\label{sec:codebleuresults}
In this study, we evaluate the semantic proximity of our synthesized repairs to the ground-truth patches using the CodeBLEU metric. Our methodology employs a CodeBLEU scoring system that quantifies similarity on a normalized scale between 0 and 1.0. This composite score accounts for lexical n-gram overlaps as well as structural alignment across Abstract Syntax Trees (ASTs) and Data-Flow Graphs (DFGs).

It is important to note that security vulnerabilities (CVEs) often permit various functionally equivalent resolutions. Thus, a CodeBLEU score below unity does not necessarily signify an unsuccessful or insecure repair. Instead, we interpret CodeBLEU as a proxy for the system's capacity to replicate the original developer's intent. To validate these semantic scores, we supplement our analysis with rigorous end-to-end compilation checks and functional testing, as discussed in Section~\ref{sec:e2e}.

\begin{figure*}[ht!] 
    \centering
    \begin{subfigure}{0.48\textwidth}
        \centering
        \includegraphics[width=\textwidth]{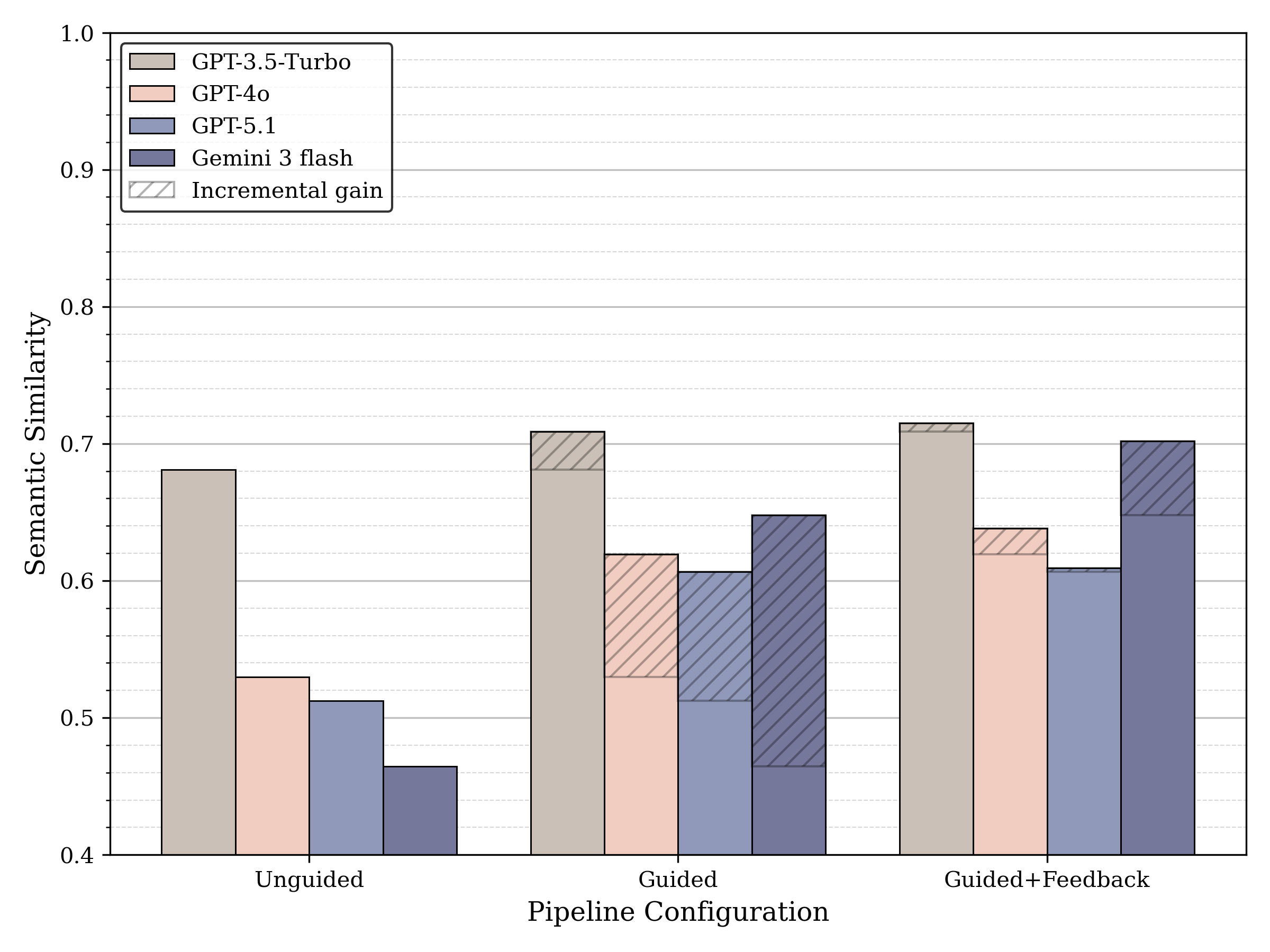}
        \caption{Python code results}
        \label{fig:codebleu_Python}
    \end{subfigure}
    \hfill 
    \begin{subfigure}{0.48\textwidth}
        \centering
        \includegraphics[width=\textwidth]{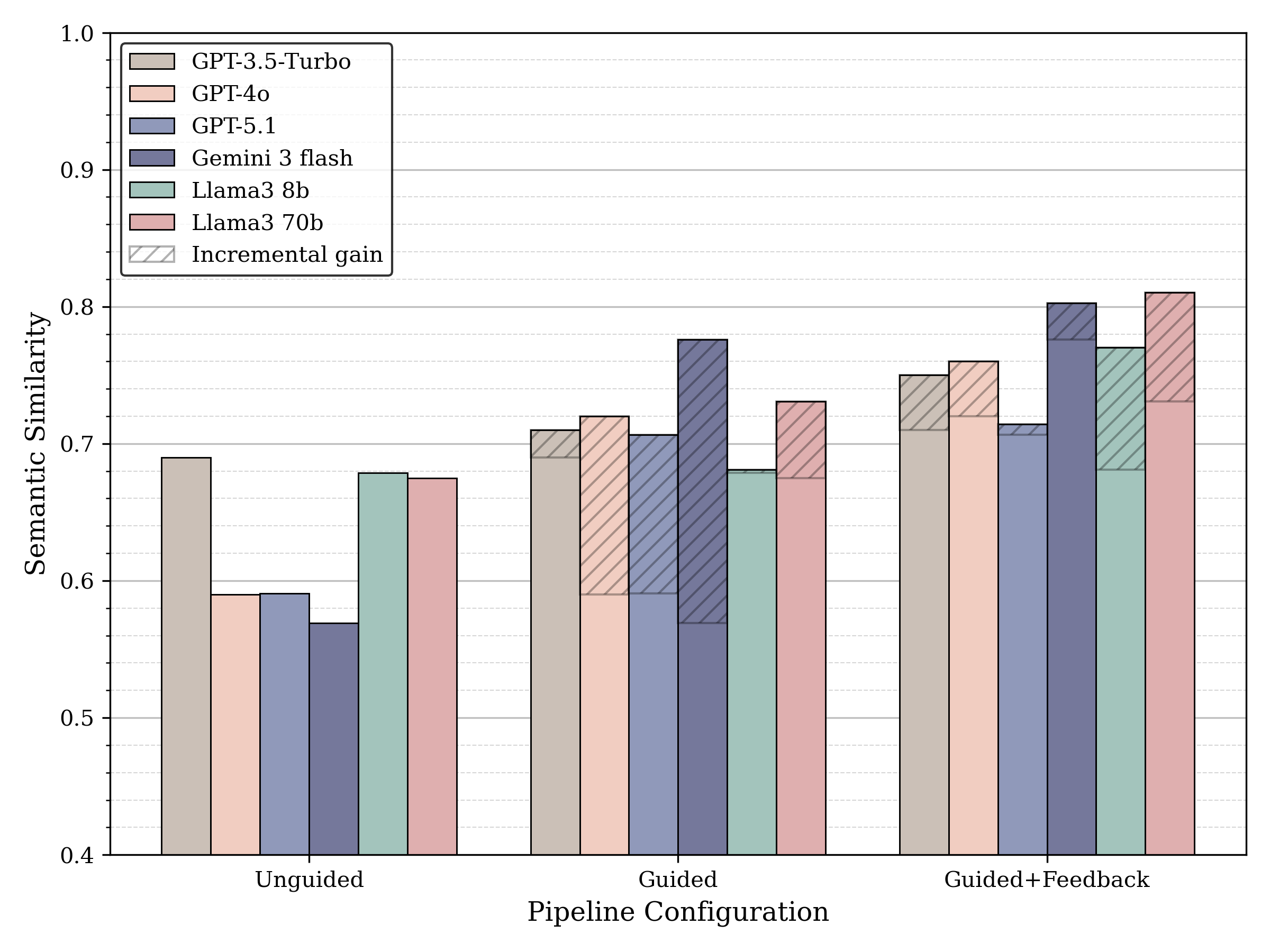}
        \caption{C code results}
        \label{fig:codebleu_C}
    \end{subfigure}
    \caption{CodeBLEU scores for all iterative pipeline configurations and LLMs. The shaded parts highlight the improvement relative to the previous configuration}
    \label{fig:codebleu}
\end{figure*}

The performance of our three experimental setups: \textit{unguided} (zero-shot), \textit{guided} (one-shot), and \textit{guided+feedback}, is illustrated in Figure~\ref{fig:codebleu}. The semantic similarity score for the fixed Python code vulnerabilities shows improvement with feedback, where Gemini 3 flash showed the biggest improvement. On the other hand, for the C language our findings indicate a substantial performance gain when utilizing the iterative feedback mechanism, with Gemini 3 flash performing the best out of the non-fine-tuned models. Additionally, the fine-tuned models demonstrated the largest improvements with the \textit{guided+feedback} configuration. Specifically, the Llama 3 70B model exhibited a 20\% improvement in CodeBLEU scores when transitioning from a one-shot approach to the \textit{guided+feedback} configuration, outperforming the GPT models and performing on par with Gemini 3 flash as shown in Figure~\ref{fig:codebleu_C}. This trend highlights the value of iterative refinement in narrowing the gap between automated patch generation and human-authored repairs.

\subsection{End-to-End Compilation}
\label{sec:e2e}

To move beyond semantic scoring and evaluate the pipeline's effectiveness in a production-grade environment, we conduct an end-to-end compilation test using real-world software. We initialize our framework with a known vulnerable function from the \texttt{iperf3} network utility, a widely used tool for TCP, UDP, and SCTP bandwidth measurement. The vulnerability, designated as \textbf{CVE-2016-4303} (associated with \textbf{CWE-120}), resides within the \texttt{cJSON\_DeleteItemFromArray} function of the integrated \texttt{cJSON} library.

Once the iterative pipeline generates a candidate patch, we assess its validity through a rigorous experimental harness. This harness simulates a malicious actor attempting to exploit the buffer overflow by providing a malformed \texttt{cJSON} object with invalid lengths. We compile two versions of the \texttt{iperf3} binary: one containing the original vulnerable code and one integrated with the LLM-generated patch. 

Upon execution, the unpatched binary suffered an immediate crash when processing the malformed object, confirming the exploit's viability. In contrast, the binary patched by our pipeline successfully identified and rejected the malicious input, thereby preventing the exploitation.

\subsection{Quantifying Engineering Effort \& Repair Latency}
\label{sec:effort}

A comprehensive evaluation of any software security solution must look beyond accuracy and consider its operational efficiency and real-world scalability. We analyze the \textit{Engineering Effort} required to secure a codebase by comparing three distinct paradigms: (1) Traditional Human Remediation, (2) Proprietary GPT-based Pipelines, and (3) Open-source LLM Pipelines augmented with LoRA.

The primary barrier to rapid patching is the reliance on skilled human labor. Trained engineers represent the highest cost and the slowest turnaround, as the remediation process requires a deep understanding of the specific codebase and cybersecurity principles. This challenge is significantly exacerbated in legacy systems, where original authors are often unavailable and documentation is missing. Industry studies estimate that the '\textit{time-to-fix}' for a single security vulnerability typically ranges from several days~\cite{mobugtime} to a full month~\cite{cvefixes}. For our comparison, we utilize a conservative estimate of 28 days for manual remediation to reflect the complexity of the CVEs in our dataset.

In contrast, our iterative pipeline demonstrates a paradigm shift in remediation latency:

\begin{itemize}
    \item \textbf{Proprietary Pipelines:} Like GPT and Gemini, they offer the lowest barrier to entry. Leveraging OpenAI’s API allows for near-instantaneous setup (approximately 1 hour) and rapid execution (5 minutes per patch) without requiring localized high-performance compute resources.
    \item \textbf{Open-source LLMs + LoRA:} Prioritize data sovereignty and long-term security. While they require a more substantial initial investment in setup time (24 hours for dataset curation and LoRA fine-tuning), they enable local, private execution that is independent of external APIs. Once trained, these models can produce candidate patches in roughly 10 minutes.
\end{itemize}

As summarized in Table~\ref{tab:timetable}, the automated approach effectively collapses the \textit{window of exposure} for vulnerable code. By combining setup and execution times into a single operational metric, we demonstrate that an iterative LLM-driven framework can substantially reduce remediation latency compared to manual workflows, suggesting significant potential value for maintaining the security of modern, high-velocity software environments.

\begin{table}[ht!]
\centering
\caption{Engineer-hours for various vulnerability patching methods.}
\scalebox{0.85}{
\begin{tabular}{ccc} 
\hline
Vulnerability Patching Technique & Setup Time      & Execution Time      \\ 
\hline
Human Intervention               & N/A        & 28 days~            \\
Open-Source LLMs+LoRAs           & 24 hours        & 10 minutes          \\
GPT LLMs                         & \textbf{1 hour} & \textbf{5 minutes}  \\
\hline
\end{tabular}
}
\label{tab:timetable}
\vspace{-0.3cm}
\end{table}

\subsection{Comparison with SOTA pipelines}
To evaluate our iterative pipeline against existing SOTA pipelines, we compare it to \textit{VulRepair}: a sequence-to-sequence model fine-tuned to fix C-code vulnerabilities. As \textit{VulRepair} requires special tags marking the vulnerable section, both pipelines are evaluated using the marked vulnerable code. Although over 62\% of \textit{VulRepair} pipeline results have extra hallucinated modifications which were not added to the final reconstructed patched code (See Sec. \ref{sec:CompareSOTAsetup}), the iterative pipeline still outperforms it across all models as shown in Table \ref{tab:VulRepair}. This demonstrates the robustness of the iterative pipeline in the vulnerability patching task.

\begin{table}[ht!]
\centering
\caption{Semantic similarity score with marked C code.}
\scalebox{.95}{
\begin{tabular}{c|cc} 
\hline
Pipeline       & Model & CodeBLEU score \\ 
\hline
\VulRepair & CodeT5 & 0.68\\
\hline
\multirow{4}{*}{Iterative}& GPT-3.5-Turbo & 0.74\\
& GPT-4o & 0.77 \\
& GPT-5.1 & 0.78 \\
& Gemini 3 flash & 0.82\\
\hline
\end{tabular}
}
\label{tab:VulRepair}

\end{table}

\section{Discussion}
\label{sec:discussion}

The generalized iterative LLM4CVE pipeline presented in this work significantly lowers the barrier to entry for repairing critical security vulnerabilities. By automating the transition from diagnosis to cure, the framework demands considerably less engineering effort, particularly for projects with diminishing knowledge bases and few active maintainers. Furthermore, the speed and efficiency of the pipeline enable near-instantaneous repair, drastically reducing the \textit{window of exposure} between bug discovery and patch deployment.

\subsection{Security Impact} 

Speeding up the vulnerability patching process has profound implications for system resilience. Our framework is especially relevant to legacy codebases, where the scarcity of domain-specific experts often leads to dangerously long turnaround times. By reducing the dependence on human intervention, the pipeline aids in fixing the critical flaws at a much faster rate, thereby increasing the overall safety of the software ecosystem. These benefits are quantitatively demonstrated in Table~\ref{tab:timetable}, which shows the drastic reduction in repair latency compared to manual efforts.

\subsection{Qualitative Patch Assessment} 

A comparative analysis between the original vulnerable function and the candidate patch generated by our pipeline reveals the depth of the model's security reasoning. To demonstrate this in practice, we provide three case studies that showcase the capabilities of the iterative pipeline across different CWEs.
\subsubsection{CWE-125: Out-of-Bounds Read}

Listing~\ref{lst:vulnpatch} illustrates the \texttt{snmp\_ber\_decode\_type} function, which is susceptible to an out-of-bounds read and potential crash due to inadequate input validation \textbf{CVE-2020-12141}. Listing~\ref{lst:gptpatch} displays the remediation synthesized by GPT-4o using the full LLM4CVE iterative framework. 

\begin{lstlisting}[label=lst:vulnpatch, caption=The original vulnerable function]
unsigned char * snmp_ber_decode_type(unsigned char *buff, uint32_t *buff_len, uint8_t *type)
{
  if(*buff_len == 0) {
    return NULL;
  }

  *type = *buff++;
  (*buff_len)--;

  return buff;
}
\end{lstlisting}

\begin{lstlisting}[label=lst:gptpatch, caption=A candidate patch generated by GPT-4o]
unsigned char* snmp_ber_decode_type(unsigned char *buff, uint32_t *buff_len, uint8_t *type) {
  if(buff == NULL || buff_len == NULL || type == NULL || *buff_len == 0) {
    return NULL;
  }

  *type = *buff++;
  (*buff_len)--;

  return buff;
}
\end{lstlisting}

Notably, while the original code attempted to validate only the length of the buffer, it did so by performing a \textit{blind de-reference} of \texttt{buff\_len} before ensuring the pointer itself was valid, a common flaw that leads to immediate segmentation faults. The iterative pipeline, through its refinement loop, successfully identified this logical oversight. The resulting patch introduces comprehensive NULL-pointer checks for \texttt{buff}, \texttt{buff\_len}, and \texttt{type} prior to any de-reference or arithmetic operations. This level of defensive programming demonstrates that the iterative process does not merely mimic human-authored patches but actively reasons about the underlying memory safety of the C language to provide a non-destructive, robust cure. This example shows the pipeline's ability to bridge the remediation gap by moving beyond simple keyword replacement toward semantic repair and rectification.

\subsubsection{CWE-119: Heap Over-Read}

The \texttt{Curl\_urldecode} function from libcurl is presented in listing~\ref{lst:vulnpatch:curl}, which is vulnerable to a heap over-read triggered by a percent sign appearing as the final character of a URL-encoded string (\textbf{CVE-2013-2174}). Listing~\ref{lst:gptpatch:curl} displays the candidate remediation synthesised by GPT-4o through the full LLM4CVE guided+feedback pipeline.

\begin{lstlisting}[label=lst:vulnpatch:curl, caption=The original vulnerable \texttt{Curl\_urldecode} function]
CURLcode Curl_urldecode(struct SessionHandle *data, const char *string, size_t length, char **ostring, size_t *olen, bool reject_ctrl)
{
  size_t alloc = (length?length:strlen(string))+1;
  /* ... */
  while(--alloc > 0) {
    in = *string;
    if(('%' == in) && ISXDIGIT(string[1]) && ISXDIGIT(string[2])) {
      /* ...... */
      string+=2; alloc-=2;
    }
    /* ... */
  }
}
\end{lstlisting}

\begin{lstlisting}[label=lst:gptpatch:curl, caption=A candidate patch generated by GPT-4o (CVE-2013-2174)]
CURLcode Curl_urldecode(struct SessionHandle *data, const char *string, size_t length, char **ostring, size_t *olen, bool reject_ctrl)
{
  size_t alloc = (length?length:strlen(string))+1;
  /* ... */
  while(--alloc > 0) {
    in = *string;
    if(('%' == in) && (alloc > 2) && ISXDIGIT(string[1]) && ISXDIGIT(string[2])) {
      /* ...... */
      string+=2; alloc-=2;
    }
    else if ('%' == in) {
      free(ns);
      return CURLE_URL_MALFORMAT;
    }
    /* ... */
  }
}
\end{lstlisting}

The vulnerability resides in the boundary condition of the percent-decoding loop in lines five to ten. As \texttt{alloc} decrements with each iteration, it reaches one when the final byte of the input buffer is being processed. Should that byte be \texttt{'\%'}, the original code unconditionally evaluates \texttt{ISXDIGIT(string[1])} and \texttt{ISXDIGIT(string[2])}, reading one and two bytes beyond the allocated region. The ground-truth fix introduced by maintainers inserts a single boundary check condition, \texttt{(alloc > 2)}, ensuring that the reads are only attempted when at least two bytes remain in the buffer. LLM4CVE pipeline independently arrives at this identical guard, confirming that the model has correctly reasoned about the relationship between the loop counter \texttt{alloc} and the lookahead indices \texttt{string[1]} and \texttt{string[2]}.

Moreover, the generated patch adds an additional layer of defense not present in the ground-truth: at lines 11 through 14 of Listing \ref{lst:gptpatch:curl}, GPT-4o adds an \texttt{else if} branch that explicitly rejects a bare \texttt{'\%'} with \texttt{CURLE\_URL\_MALFORMAT}, preventing the character from being forwarded to the caller as literal data. This augmentation, absent from the official commit, demonstrates that the iterative pipeline does not merely replicate the minimal human patch but can synthesize complementary defensive measures through its self-refinement loop.


\subsubsection{CWE-787: Stack Buffer Overflow}

The vulnerable and LLM4CVE generated patched versions of the \texttt{add\_password} function from the \texttt{pam\_radius} PAM authentication module are shown in Listing~\ref{lst:vulnpatch:pam} and Listing~\ref{lst:gptpatch:pam} respectively. This function is susceptible to a stack-based buffer overflow \textbf{CVE-2015-9542}. 

\begin{lstlisting}[label=lst:vulnpatch:pam, caption=The original vulnerable \texttt{add\_password} function]
static void add_password(AUTH_HDR *request, unsigned char type, CONST char *password, char *secret)
{
  int length = strlen(password);
  unsigned char hashed[256 + AUTH_PASS_LEN];
  if (length > MAXPASS) {
    length = MAXPASS;
  }
  /* ... */
  /* round length up to nearest 16-byte block */
  /* ... */
  memset(hashed, 0, length);
  memcpy(hashed, password, strlen(password));
  /* ... */
}
\end{lstlisting}

\begin{lstlisting}[label=lst:gptpatch:pam, caption=A candidate patch generated by GPT-4o (CVE-2015-9542)]
static void add_password(AUTH_HDR *request, unsigned char type, CONST char *password, char *secret)
{
  int length = strlen(password);
  unsigned char hashed[256 + AUTH_PASS_LEN];
  if (length > MAXPASS) {
    length = MAXPASS;
  }
  /* ... */
  /* round length up to nearest 16-byte block */
  /* ... */
  memset(hashed, 0, length);
  memcpy(hashed, password, length);
  /* ... */
}
\end{lstlisting}

While the function correctly constrains \texttt{length} to \texttt{MAXPASS}, the subsequent \texttt{memcpy} in line 12 ignores this sanitized value entirely, instead passing the raw length of the caller supplied password \texttt{strlen(password)}. An attacker submitting a password exceeding \texttt{length} can therefore overflow \texttt{hashed}, corrupting the stack frame, potentially manipulating saved registers' states or even the function return address.

The LLM4CVE candidate patch, generated under the guided+feedback configuration, produces the ground truth solution, replacing \texttt{strlen(password)} with the bounded variable \texttt{length}. This repair is identical to the ground-truth CVEfixes commit. The precision of this result is significant: reaching it requires not merely recognising the class of vulnerability (unbounded copy into a fixed buffer), but identifying that the correct bound had already been computed earlier in the same scope and was available under the name \texttt{length}.

\subsection{The Performance Gradient: GPT/Gemini vs. Llama}

A central aim of this study is to investigate the \textit{performance delta} between high-capacity proprietary models and optimized open-source architectures. Throughout the experimental phase, the \textbf{Llama 3 70B} model demonstrated consistent parity with the other evaluated LLMs, most notably when deployed within the integrated \textit{Guided + Feedback} configuration of the LLM4CVE framework.

The substantial advancements in generative efficacy were largely a byproduct of our pipeline's iterative structure, which we attribute to the efficacy of domain-specific adaptation on open-source weights. While proprietary systems like \textbf{GPT-5.1} enable significant general-purpose reasoning capabilities, their \textit{black-box} architecture constrains the implementation of the granular LoRA strategies described in Section~\ref{sec:methods}. Consequently, our results suggest that for specialized vulnerability fixing tasks, medium-scale, adaptable open-source models such as Llama 3 70B offer a more resilient framework than the reliance on general-purpose proprietary APIs.

\section{Limitations and Future Work}
\label{sec:limits}

While LLM4CVE demonstrates a significant leap in bridging the remediation gap, several technical and structural limitations remain that define the boundaries of current LLM-based program repair. Addressing these constraints is necessary for the evolution of fully autonomous security agents capable of handling the complexities of modern software engineering.

\subsection{Architectural and Contextual Constraints}

One primary limitation of the current pipeline is its focus on \textit{intra-procedure repair}. While LLM4CVE excels at fixing vulnerabilities contained within a single function, such as buffer overflows or NULL-pointer dereferences, many real-world security flaws are \textit{interprocedural} in nature. Vulnerabilities like race conditions, complex logic errors, or broken authentication flows often require coordinated changes across multiple files or modules. Current context window constraints and the quadratic complexity of standard attention mechanisms limit the model's ability to maintain a global view of large-scale repositories. This can potentially lead to \textit{fragmented patches} that fix a local symptom while ignoring the systemic cause or inadvertently breaking dependencies in distant parts of the codebase.

\subsection{Semantic Hallucinations and Symbolic Guardrails}
Despite achieving a high quality score, LLMs remain susceptible to \textit{semantic hallucinations}: the generation of code that appears syntactically correct but is logically insecure or non-functional. In security contexts, this may manifest as the introduction of secondary vulnerabilities while fixing an existing one. While our evaluation did not explicitly quantify the rate of secondary vulnerability introduction, this remains a recognized risk in security-focused code generation. Current research suggests that high-autonomy agentic workflows can occasionally introduce subtle logic flaws if not constrained by symbolic guardrails. Our reliance on CodeBLEU, while effective for semantic similarity, does not provide the mathematical guarantees required for mission-critical software.

\subsection{Future Research Directions}
To overcome these barriers, future iterations of LLM4CVE should pivot toward an \textit{Agentic and Symbolic Hybrid} architecture. We identify four key pillars for the continued development of this framework:

\begin{itemize}

\item \textbf{Dynamic Trace Integration:} Moving beyond static analysis, future versions of LLM4CVE can integrate execution-level dynamic information, such as variable states and call stacks, into the feedback loop. This would allow the model to simulate runtime behavior and significantly reduce the patch search space by identifying precisely which logical branches lead to a crash.

\item \textbf{Formal Verification Oracles:} To eliminate the risk of hallucination and secondary vulnerability introduction, future efforts on LLM4CVE can incorporate formal verification methods, such as SMT solvers, as \textit{truth oracles} within the pipeline. By checking patches against formal behavioral specifications, the system can provide mathematical proofs of security that purely probabilistic models cannot capture.

\item \textbf{Multi-Agent Orchestration:} Next works on LLM4CVE can explore the use of multi-agent ecosystems where specialized \textit{security reviewer} agents critique the patches generated by \textit{repair} agents. This adversarial collaboration mimics the human peer-review process, potentially raising the reliability of automated software code repair to near-perfect standards.

\item \textbf{Cost-Performance Optimization:} Future work can investigate the empirical trade-offs between execution costs and repair performance. Specifically, varying the number of iterations within the LLM4CVE pipeline will help determine the point of diminishing returns, optimizing the balance between computational overhead and patch success rates.

\end{itemize}

By addressing these limitations, LLM4CVE will evolve from a specialized assistant into a true \textit{coding collaborator and partner}, capable of securing the software supply chain at a scale and speed unattainable by human experts alone.

\section{Conclusion}
\label{sec:conclusion}

The persistence of the \textit{Found, but Not Fixed} phenomenon in software security underlines a critical bottleneck in modern vulnerability management. In this work, we introduced LLM4CVE, a generalized, iterative pipeline designed to autonomously remediate critical software vulnerabilities with minimal human intervention. By bridging the gap between conventional bug reporting and advanced LLM reasoning, our framework transforms static diagnostic data from CVE and CWE repositories into dynamic code patching agents. This transition represents a paradigm shift in how we handle legacy software security, moving from reactive documentation to proactive, autonomous cure.

Our findings offer several significant insights for the software security community, most notably the power of iteration over the standard one-shot generation paradigm. Our results demonstrate that the introduction of a closed-loop feedback mechanism, informed by CodeBLEU semantic analysis and compiler logs, yielded a 20\% improvement in patch fidelity. This suggests that future APR tools must prioritize agentic, recursive reasoning to handle complex security logic effectively. Furthermore, we observed that specialized, open-source models like Llama 3 70B, when augmented with LoRA, can match the performance of much larger proprietary \textit{black-box} models with the closed-loop feedback. This highlights a viable path for organizations to deploy on-site, privacy-preserving remediation tools without sacrificing accuracy or technical depth.

The framework effectively collapses the \textit{window of exposure} for code vulnerabilities, providing a scalable method for maintaining system integrity in high-velocity software environments. As we look toward the future of expert-free program repair, extending this pipeline to multi-file structural changes and integrating dynamic execution traces will be essential next steps in the evolution of fully autonomous vulnerability remediation.

\bibliographystyle{elsarticle-num}
\bibliography{example}

\end{document}